# Electron tunneling spectroscopy of a quantum antidot in the quantum Hall regime

V. J. Goldman
*Department of Physics, Stony Brook University, Stony Brook NY 11794-3800, USA*
Jun Liu and A. Zaslavsky
*Department of Physics and Division of Engineering, Brown University, Providence RI 02912*

Quantum antidot, a small potential hill introduced into a two-dimensional electron system, presents an attractive tool to study quantum mechanics of interacting electrons. Here we present experiments on electron resonant tunneling via a quantum antidot on the integer $i$ = 1, 2, 3, 4, 5, and 6 quantum Hall plateaus. Several new features are reported. First, as a function of magnetic field, we observe up to six quasiperiodic resonant tunneling peaks within the fundamental flux period: when flux $h/e$ is added to the area of the antidot there are $i$ peaks on the $i$-th integer plateau, when $i$ spin-polarized Landau levels are occupied. Corresponding back gate voltage data show one peak per added charge $e$ on all integer plateaus. Second, we observe tunneling dips in four-terminal resistance ("forward scattering") on the even $i$ = 2, 4, and 6 plateaus, when population of both spins is nearly equal. We also report an internal structure observed within the $h/e$ period: on the $i$ = 3 spin-split plateau, two of the three resonant tunneling peaks are higher and/or closer than the third. Puzzlingly, in this regime, when the back gate voltage is swept, the tunneling peaks are grouped in pairs. These results are attributed to the dominance of the electron-electron Coulomb interaction, effectively mixing Landau level occupation, and to the self-consistent electrostatics of the antidot.

## I. INTRODUCTION

The integer quantum Hall effect[1] can be understood in terms of electron transport by edge channels corresponding to an integer number of fully occupied Landau levels.[2,3] Near an integral Landau level filling $\nu \approx i$, the Hall resistance is quantized exactly to $h/ie^2$ because the chemical potential lies in the gap of localized bulk states, and the current is carried by dissipationless edge channels. Dissipative transport occurs when current is carried either by extended bulk states of the partially occupied topmost Landau level, between the plateaus, or by quantum tunneling between the extended edge states. Such interpretation of the integer quantum Hall effect in terms of edge channels is straightforward for non-interacting electrons, when the edge channels are formed in one-to-one correspondence with the bulk Landau levels defined in the single-electron density of states.[2,3] However, under nearly all experimental conditions, the electron-electron interaction is not small compared to single-particle energies involved, and the effects of interaction are subjects of intense experimental and theoretical research.

Quantum antidots (QAD) present a fascinating tool to study fundamental many body quantum mechanics. For example, a QAD electrometer has been used in the direct experimental observation of a fractionally quantized electric charge of Laughlin quasiparticles.[4-6] Other earlier studies of antidots performed in the integer quantum Hall regime focused on demonstration of the Aharonov-Bohm effect in the edge channel circling the antidot;[7-9] detection of the variation of the charge state of the antidot when one electron is added or subtracted;[10] and a detailed study of the line shape of a single resonant tunneling peak and its temperature dependence in the limit where the tunneling peaks are well separated.[11] QADs were also studied in the fractional quantum Hall regime, in particular the line shape of the tunneling peaks has been studied in detail,[12] and the absolute energy scale of the QAD-bound states was determined via the technique of thermal activation.[13] An experimental observation of a coherent QAD "molecule"[14] has led to a proposal of an anyonic quantum computation scheme based on adiabatic transfer of Laughlin quasiparticles in arrays of coupled QADs.[15]

In this paper we report several new features observed in resonant tunneling experiments through a single QAD placed in a constriction in the integer quantum Hall regime. First, we observe up to six quasiperiodic tunneling peaks when flux $h/e$ is added to the area of the antidot: $i$ peaks on the $i$-th integer plateau, when $i$ spin-split Landau levels are occupied. Second, we report observation of tunneling dips in four-terminal resistance ("forward scattering") on the even $i = 2$, 4, and 6 plateaus, when both spin-up and spin-down electron populations are nearly equal. We also report internal structure within the fundamental $h/e$ period: on the $i = 3$ spin-split plateau two of the three resonant tunneling peaks are higher and/or closer than the third. Puzzlingly, the periodicity is different when back gate voltage is swept: the tunneling peaks group in pairs.

## II. AN ISOLATED ANTIDOT

In this Section we outline non-interacting electron theory of an isolated quantum antidot and discuss its inadequacy under experimental conditions. An antidot is created when a small potential hill $U(r)$ is introduced into a two-dimensional electron system (2DES) in the presence of quantizing magnetic field, Fig. 1. As is well known two-dimensional electrons in a strong perpendicular $\mathbf{B} = B\hat{\mathbf{z}}$ form Landau levels.[2,3,16] The Hamiltonian can be written as

$$H = \sum_j \left\{ \frac{1}{2m^*}[\mathbf{p}_j + e\mathbf{A}(\mathbf{r}_j)]^2 - \mu_B g^* \mathbf{S}_j \cdot \mathbf{B} - eU(\mathbf{r}_j) \right\} + \sum_{j<k} \frac{e^2}{4\pi\varepsilon\varepsilon_0 |\mathbf{r}_j - \mathbf{r}_k|}, \qquad (1)$$

where $j$-th electron with charge $-e$, effective GaAs conduction band mass $m^* = 0.067 m_e$ and spin Lande factor $g^* = -0.44$ experiences vector potential $\mathbf{A}(\mathbf{r}_j) = \tfrac{1}{2}(\mathbf{B} \times \mathbf{r}_j)$ (in the symmetric gauge) and the antidot bare potential $U(r_j)$, which is assumed to be rotationally-symmetric. The two-dimensional electrons are described by radius-vector $\mathbf{r}_j$, momentum $\mathbf{p}_j$ and spin $\mathbf{S}_j$ operators. The double sum gives the contribution of the interelectron Coulomb interaction in a host medium with dielectric constant $\varepsilon$.

First, neglecting electron-electron interaction and the antidot bare potential $U(r)$, in the symmetric gauge, single-particle orbitals $\psi_m$ in each Landau level can be chosen to be eigenstates of the angular momentum operator $\mathbf{L} = \mathbf{r} \times \mathbf{p}$ with eigenvalues $\hbar m$, where quantum numbers $m = 0, 1, 2, \ldots$. For an electron in the lowest Landau level (Landau level index $N = 0$) these orbitals are

$$\psi_m(r, \vartheta) = r^m \exp(im\vartheta - r^2/4) / \sqrt{2\pi 2^m m!}, \qquad (2)$$

where $r$ is in units of magnetic length $\ell = \sqrt{\hbar/eB}$, and $\vartheta$ is the azimuthal angle. Analogous basis wave functions $\psi_{m,N,\pm}$ can be written for all Landau levels. In each spin-polarized Landau level all eigenenergies are equal: $E_{m,N,\pm} = \hbar\omega_C(N + 1/2) \mp (1/2)\mu_B g B$ does not depend on $m$. That is, the states $\psi_{m,N,\pm}$ are all degenerate for a given $N$ and spin. As is easy to see, for $m \gg 1$ the probability density $|\psi_m|^2$ is sharply peaked at $r = r_m = \sqrt{2m}\ell$, and the area within a circle of radius $r_m$ is $S_m = 2\pi m \ell^2 = mh/eB$. In other words, the semiclassical area of the orbital $\psi_{m,N,\pm}$ in each Landau level encloses precisely $m(h/e)$ of magnetic flux, independent of $N$ and spin, known as the Aharonov-Bohm quantization condition.[4-6]

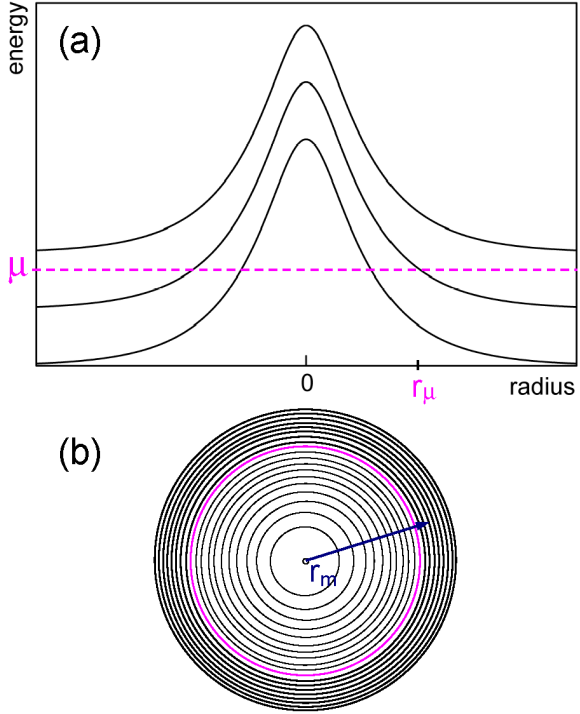

FIG. 1. An isolated quantum antidot. (a) Neglecting electron-electron interaction, the three lowest Landau levels shown follow the bare potential $U(r)$ hill. At a low temperature, the chemical potential $\mu$ separates the occupied and empty electron states. (b) A view of QAD-bound electron orbitals in the 2D electron plane for a weak, rotationally-symmetric $U(r)$. The $m$-th orbital of radius $r_m$ encloses $mh/e$ of magnetic flux. The occupied states ($r_m > r_\mu$) are shown by thicker circles.

The main effect of the antidot bare potential $U(r)$ is to lift the massive degeneracy of the antidot-bound electron states $\psi_m$ in each Landau level, Fig. 1. When the gradient of $U(r)$ is small, it can be treated as a perturbation. The condition of weakness of $U(r)$ is that the potential energy acquired by an electron displaced from $r_{m+1}$ to $r_m$ be small compared to the energy separating successive Landau levels: $(\partial U/\partial r)(r_{m+1} - r_m) \ll \hbar\omega_C = \hbar eB/m^*$, so that the external potential does not induce significant Landau level mixing. We note that $r_{m+1} - r_m = \ell^2/r_m = \ell/2m \propto \sqrt{B}/B = 1/\sqrt{B}$ on a given quantum Hall plateau $\nu \approx i$.

In the first order of the perturbation theory each energy $E_{m,N,\pm}$ is shifted by $\langle \psi_{m,N,\pm} | U(r) | \psi_{m,N,\pm} \rangle \cong U(r_m)$, independent of Landau level index and spin. Thus, to the first order, the non-interacting QAD-bound electron energy is

$$E_{m,N,\pm} \cong (N+1/2)\hbar\omega_C \mp (1/2)\mu_B g^* B + U(r_m). \qquad (3)$$

The lifting of degeneracy by $U(r)$ allows to change population of the antidot-bound states one particle at a time by tuning an external parameter, such as magnetic field or a gate voltage, provided the temperature, electromagnetic background "noise", and any applied excitation is low enough. This, in turn, allows tunneling spectroscopy of the antidot-bound electron states, the subject of this work.

The second order contribution[17] to $E_{m,N,\pm}$ involves mixing of Landau levels of the same spin and, for $m \gg 1$, is approximately $-[(\partial U/\partial r)\ell]^2/4m\hbar\omega_C$. Linearizing $U(r)$ we obtain the condition

$$U_{max} \ll 2\hbar\omega_C (R_{QAD}\sqrt{2m}/\ell) \propto B^{3/2}, \qquad (4)$$

where $U_{max}$ is the height of the QAD potential hill and $R_{QAD}$ is the characteristic radial size of the bare potential. This condition is nearly always satisfied in experiments because in GaAs, taking $R_{QAD} = 300$ nm (typical depth of the 2DES layer from the surface) we obtain $U_{max} \ll 0.64$ eV at 1 Tesla, $U_{max} \ll 1.8$ eV at 2 Tesla and $U_{max} \ll 9.4$ eV at 6 Tesla.[10,11] Since the interacting electrons screen the bare potential, the screened value of $\partial U/\partial r$ at the chemical potential should be used above, further relaxing the condition on $U_{max}$, Eq. (4).

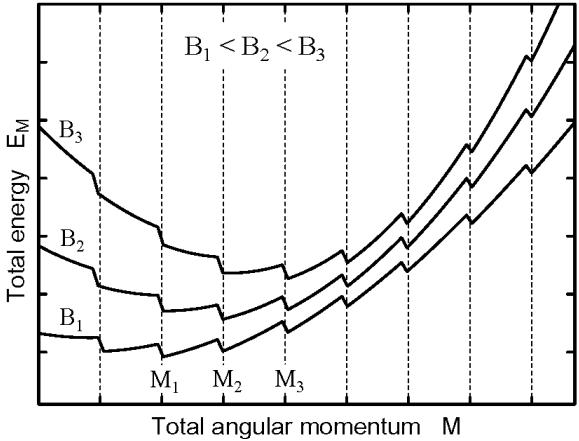

FIG. 2. Qualitative dependence of total energy $E_M$ vs. total angular momentum $M\hbar$ for interacting QAD-bound electrons at three applied magnetic fields. The ground state of the electron system shifts from $M_1$ at $B_1$ to $M_2$ at $B_2$, and to $M_3$ at $B_3$.

The effect of the electron-electron interaction is to mix the occupation of the basis orbitals $\psi_{m,N,\pm}$ belonging both to the same spin-polarized Landau level, and to different Landau levels, so that the Landau level index $N$ is no longer a good quantum number. The many-electron ground states of definite spin (+ or −) and the total angular momentum $M\hbar$ are constructed as

$$\Psi_{M,\pm} = \sum_{m,N} c_{m,N,\pm} \psi_{m,N,\pm}. \qquad (5)$$

These $\Psi_{M,\pm}$ involve superposition of a number of the basis orbitals in a number of Landau levels of a given spin.

When electron-electron interaction dominates, an analogy with the fractional quantum Hall effect[16,18] suggests total energies of QAD-bound electrons $E_{M,\pm}$ behave as illustrated in Fig. 2. Energies $E_{M,\pm}$ exhibit a cusp down at integer values of $M$, and one of these cusp-down values is the global ground state of the system at a given $B$. When magnetic field is increased, at some $B$ the energies $E_{M,\pm}$ and $E_{M+1,\pm}$ cross, and the total angular momentum of the ground state is incremented by one. When $E_{M,\pm}$ and $E_{M+1,\pm}$ are equal, it costs no energy to add an electron to the system at the chemical potential, so that electrons at $\mu$ can tunnel resonantly between the two edges via the QAD and a conductance peak occurs. This is similar to tunneling dynamics in Coulomb-blockade systems.

Gauge invariance arguments[16,19] require that when fluxoid $\Delta_\Phi = h/e$ is inserted adiabatically at the center of the antidot (where there are no electrons), the electron system returns to the initial microscopic state. Thus, for quantum antidots $\Delta_\Phi = h/e$ is the fundamental flux period. As discussed in Sect. V, on $i$-th quantum Hall plateau addition of flux $h/e$ increments $M$ by $i$ to $M+i$, so that there are $i$ tunneling peaks expected within the fundamental flux period. For large $M \gg 1$, when $r_\mu$ is nearly fixed by the self-consistent confinement potential, the corresponding field interval is $\Delta_B = h/e\pi r_\mu^2$.

In 2D electron samples realized in GaAs/AlGaAs heterostructures, for electron density $n = 1 \times 10^{11}$ cm$^{-2}$, at 1 Tesla the characteristic energies are: cyclotron $\hbar\omega_C = 1.7$ meV, Zeeman $\mu_B g^* B = 0.025$ meV, while interelectron Coulomb interaction $e^2 n^{1/2}/4\pi\varepsilon\varepsilon_0 = 3.6$ meV dominates. Hartree-Fock calculations[20] for up to 300 electrons forming an $i=2$ "maximum density droplet", and a density functional calculations[21] in antidot geometry demonstrate some of the qualitative features of the interacting 2D electrons discussed above.

### III. EXPERIMENTAL TECHNIQUE

The quantum antidot samples were fabricated from a very-low disorder GaAs/AlGaAs heterojunction material. The 2D electron layer (320 nm below the surface) with "bulk" electron density $n_B = 1.2 \times 10^{11}$ cm$^{-2}$ is prepared by exposure to red light at 4.2 K. The two independently-contacted front gates and the antidot were defined by electron beam lithography on a pre-etched mesa with Ohmic contacts. After a shallow, 150 - 180 nm wet chemical etching, 50 nm thick Au/Ti gate metal was deposited in etch trenches, followed by lift-off. Samples were mounted on sapphire substrates with In metal, extending over the entire GaAs chip, which serves as a global backgate. The QAD sample reported in this paper has nominal lithographic antidot diameter of 180 nm and the antidot-front gate distance of 750 nm. It was measured in several cooldown cycles over three years; during this time surface depletion of the etched GaAs was affected to some extent by oxidation in room air, but the pertinent tunneling and transport features reported here were observed to persist.

Samples were cooled to 12 mK in the mixing chamber tail of a top-loading into mixture $^3$He-$^4$He dilution refrigerator. Four-terminal resistance $R_{XX} \equiv V_X/I_X$ was measured by passing a 100 to 500 pA (larger current on $i > 1$ plateaus), 5.4 Hz AC current through contacts 1 and 4, and detecting the voltage between contacts 2 and 3 (see Fig. 3) by a lock-in-phase technique. An extensive cold filtering cuts the integrated electromagnetic "noise" environment incident on the sample to $\sim 5 \times 10^{-17}$ W, which allows us to achieve a very low electron temperature of 18 mK in a mesoscopic sample.[12]

The antidot and the two front gates are deposited into etch trenches. Even when voltages applied to the front gates $V_{FG} = 0$, the GaAs surface depletion potential of the etch trenches defines two constrictions in 2DES, separating the front gates from the circular antidot. In this work, the $V_{FG}$ (with respect to the 2D electron layer) are approximately equal; the difference is used to fine tune for the symmetry of the two constrictions. The depletion potential has a saddle point in each constriction region, and so has the resulting electron density profile. From the magnetotransport measurements (see Section IV), we estimate the saddle point density value $n_C \approx 0.9 n_B$ when $V_{FG} = 0$, which varies somewhat due to the self-consistent electrostatics of the 2D electrons in presence of a quantizing magnetic field. Upon application of a negative $V_{FG} \approx -1.3$ V, the constriction saddle point density is reduced to $n_C \approx 0.58 n_B$.

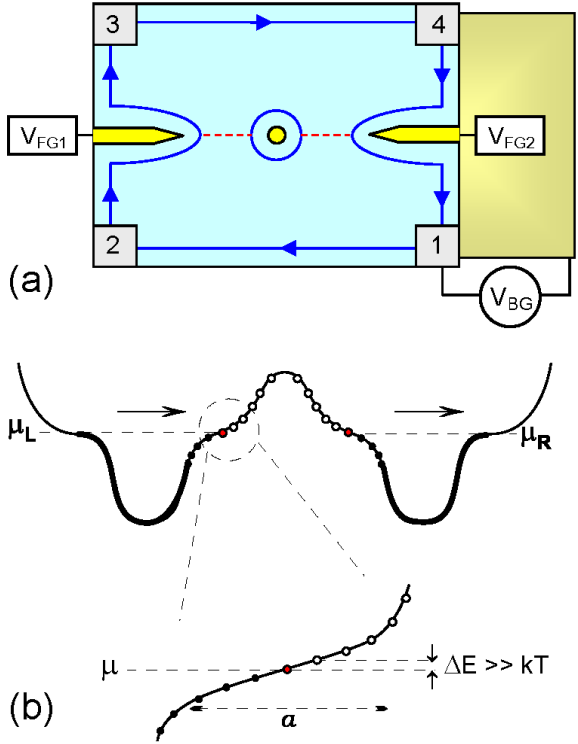

FIG. 3. A quantum antidot sample. (a) The antidot is in the constriction between two front gates (FG). Numbered rectangles are Ohmic contacts; the blue arrowed lines show an edge channel. The red dashed lines show the resonant tunneling path. The back gate (BG) extends over the entire sample on the opposite side of the insulating GaAs substrate. (b) Self-consistent energy diagram of one Landau level in the constriction. The energy spectrum is continuous at the extended edges and discrete at the antidot. The arrows show tunneling at chemical potential $\mu$; the quantum Hall gap forms the tunneling barriers. At a low temperature, tunneling between the left and right edges occurs via only one antidot-bound state within $kT$ of $\mu$.

## IV. QUANTUM HALL MAGNETOTRANSPORT

The 2D electron system on a quantum Hall plateau $i$ opens an energy gap and therefore is an insulator. The quantum Hall edge channels are formed following the equipotentials, where the electron local density $n$ is such that the Landau level filling factor $\nu = hn/eB$ is equal to an integer $i = 1, 2, 3, \ldots$. While $\nu \propto n/B$ is a variable, the quantum Hall exact filling $i$, defined as the inverse of value of the *quantized* Hall resistance $R_{XY}$ in units of $h/e^2$ (that is, $i \equiv h/e^2 R_{XY}$), is a quantum number. In the 2D bulk, variation of $B$ from the exact filling $\nu = i$ is accommodated by creation of quasiparticles ($\nu > i$) or quasiholes ($\nu < i$). Here the relevant quasiparticles are electrons of charge $e$ and Fermi statistics in the next ($\nu = i+1$) partially occupied Landau level, and quasiholes are the missing electrons in the otherwise full $\nu = i$ Landau level.

The edge channels on the periphery of the 2DES have a continuous energy spectrum because they are macroscopically long and are connected to a dephasing electron reservoir, an Ohmic contact. The transport current is carried by the extended states in the edge channels near the chemical potential, where low-energy excitation is possible. The particle states of the edge channel circling the antidot are of a microscopic size, and, if quantum-coherent when the temperature and excitation are sufficiently low, are quantized by the Aharonov-Bohm condition, as discussed in Section II. Because of the finite gradient of the antidot potential $U(r)$, the QAD-bound electron states have a non-degenerate energy spectrum. Resonant tunneling between the extended edge channels proceeds via the quantized antidot-bound states. On a plateau, when the bulk is gapped, the tunneling is the only transport mechanism giving rise to non-zero $R_{XX}$, the quantum Hall gap forms the tunneling barrier.

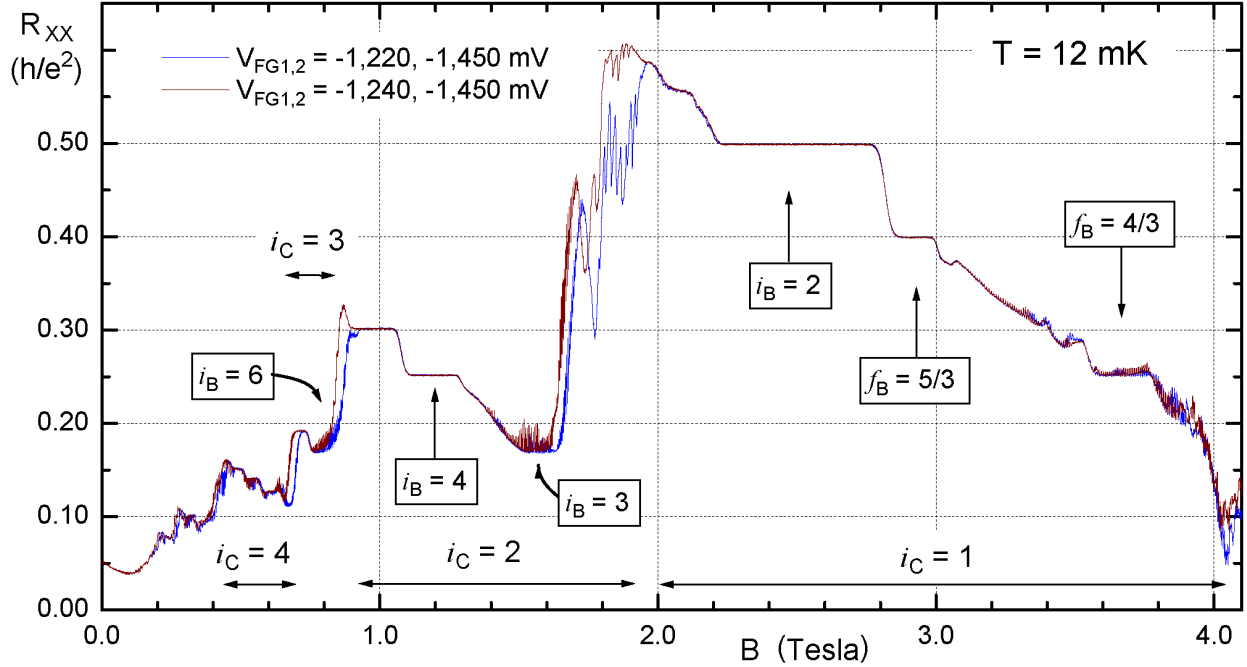

FIG. 4. Four terminal $R_{XX}$ vs $B$. The two traces were obtained with two slightly different biases on one of the front gates, $V_{FG1}$. Note that the bulk filling does not depend on front gate bias.

Figure 4 shows the directly measured four-terminal $R_{XX}$ as a function of applied magnetic field $B$. The two features seen are the quantized $R_{XX}$ plateaus, discussed below, and the resonant tunneling regions, presented in detail in the next Section. The *local* Landau level filling factor $\nu = hn/eB$ is proportional to $n(r)$, and the electron density in the constrictions $n_C$ is appreciably less than $n_B$ in the bulk. There are two regimes possible: one when the two quantum Hall plateaus, $i_C$ in the constrictions and $i_B$ in the 2D bulk, overlap in a range of $B$, resulting in a *quantized* $R_{XX} = (h/e^2)(1/i_C - 1/i_B)$ plateau.[22,23] Several such examples are seen in Fig. 4. Increasing magnetic field, a transition $i_B \to i_B - 1$ is seen as a step down, $R_{XX}(B)$ decreases, and a transition $i_C \to i_C - 1$ is seen as a step up, $R_{XX}(B)$ increases fast. This also can occur for a fractional quantum Hall plateau, for example $i_C = 1$, $f_B = 5/3$ and $f_B = 4/3$ in Fig. 4. The second possibility is when $\nu_B \approx 1.5$ or $2.5$ occurs on a well-developed $i_C = 1$ or $i_C = 2$ plateau. Here the bulk $R_{XY}(B) \approx h/\nu e^2$ is approximately linear, and is seen as a negative slope straight line in the four-terminal $R_{XX}(B)$, $B \approx 3.2$ T ($\nu_B \approx 1.5$, $i_C = 1$) and $B \approx 1.4$ T ($\nu_B \approx 3.5$, $i_C = 2$). Thus observation of a quantized plateau in $R_{XX}(B)$ implies quantum Hall plateaus for both the constriction region and the bulk, and in practice provides definitive values for both $i_C$ and $i_B$.

## V. RESONANT TUNNELING

When the constriction is on a quantum Hall plateau no dissipative conduction is possible between the right and the left edges, conductance $G = 0$ in the limit of low temperature and excitation, except that the electrons can tunnel resonantly via the QAD-bound states, giving rise to quasiperiodic tunneling conductance $G_T$ peaks. A peak in $G_T$ occurs when a QAD-bound state crosses the chemical potential, see Fig. 3, and therefore signifies the change in QAD occupation by one electron. The directly measured $R_{XX}$ vs $B$ data on constriction $i_C = 1$ to 6 plateaus, obtained at 12 mK, is shown in Figs. 5 – 10.

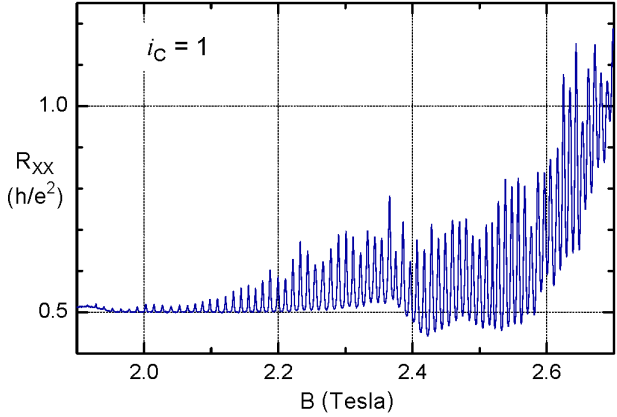

FIG. 5. Resonant tunneling data for $i = 1$ plateau in the QAD constriction.

Near the center of the plateau the tunneling is weak, $G_T << e^2/h$. The amplitude of the $R_{XX}$ peaks on a given quantum Hall plateau is expected to increase monotonically with decreasing $\nu$ (increasing $B$ or decreasing $V_{BG}$), because $r_\mu$ increases and also the front gate edge channels move closer to the antidot, see Fig. 3, so that the tunneling distance decreases. For the same reason, the amplitude of the $R_{XX}$ dips increases with increasing $\nu$. In addition, smooth non-monotonic modulation of the peak amplitude has been attributed to "mesoscopic effects",[8,9] such as modulation of the tunneling amplitude by the residual disorder potential. Another interesting feature seen in some data is likely due to mesoscopic effects: for example, in the $i = 4$ upper trace of Fig. 8, every seventh peak is smaller than its neighbors; such behavior, however, is sensitive to small variation of front-gate voltage. In contrast, the $i = 3$ subperiodic structure is more robust; it is discussed further in the next Section.

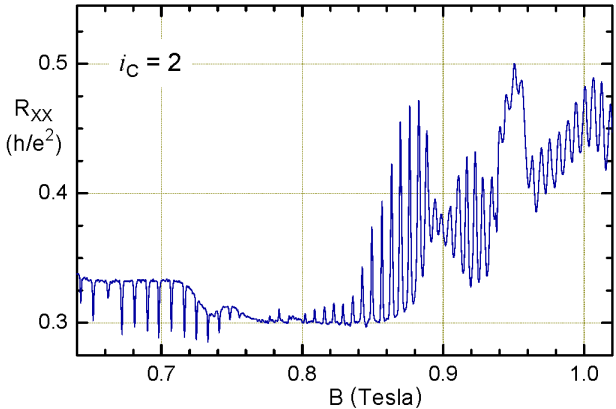

FIG. 6. Resonant tunneling data for $i = 2$ in the constriction. The $R_{XX}$ peaks are seen at constriction filling $\nu_C < 2$, and the $R_{XX}$ dips for $\nu_C > 2$.

The $R_{XX}$ dips are observed on the low-$B$ side ($\nu > i$) of the even-filling plateaus, Figs, 6, 8, and 10. These dips are often attributed to "forward scattering", that is, to tunneling across the constriction, perpendicular to the "back scattering" direction shown in Fig. 3(a). Since the filling in the bulk is greater than in the constriction, it is possible to have two edge channels between the constriction and the bulk, on either side of the antidot-containing constriction. The forward scattering is visualized as proceeding via the two tunneling links, coupling the two bulk edge channels, on either side of the antidot. Provided the two tunneling amplitudes are nearly equal, resonant tunneling conductance peaks will result. For forward scattering, edge-network models predict the four-terminal $R_{XX}$ dipping by $\approx G_T/(ie^2/h)^2$ below the quantized plateau value.

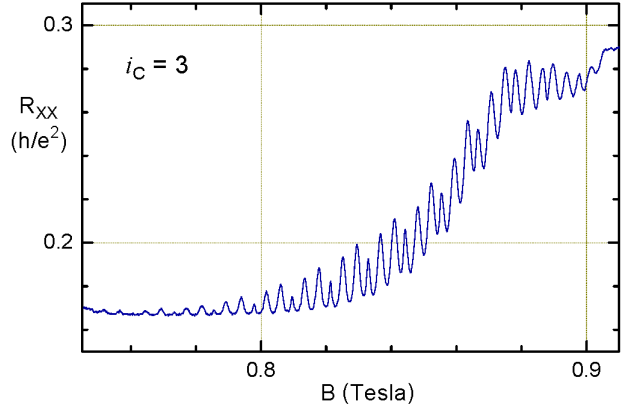

FIG. 7. Resonant tunneling data for $i = 3$ in the QAD constriction. Note the internal structure within the fundamental flux period, which contains three tunneling peaks.

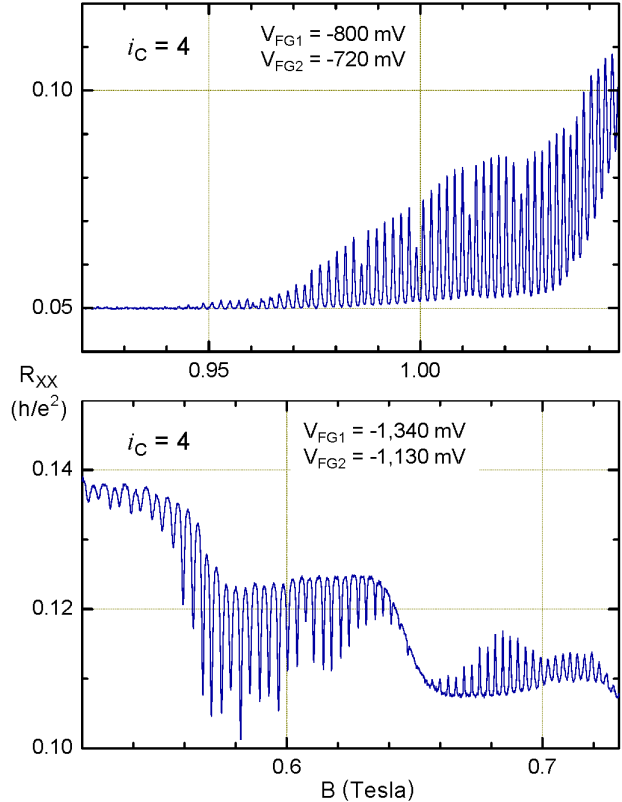

FIG. 8. Resonant tunneling data for $i = 4$ in the QAD constriction. The $R_{XX}$ peaks are seen at $\nu_C < 4$, and the $R_{XX}$ dips for $\nu_C > 4$.

However, several experimental aspects are puzzling if such resonant forward scattering is the origin of the $R_{XX}$ dips. First, the magnetic field period of the dips is 20 – 30% greater than that for the peaks on the same plateau. This indicates a noticeably smaller area associated with the dips than with the peaks, which is hard to reconcile with the forward scattering edge path having to enclose a larger area in the antidot geometry. Second, the imperfection of the antidot lithography has to be compensated for by tuning the front-gate bias so as to achieve resonance in the two tunneling amplitudes. It is very difficult to believe that both forward- and backscattering amplitudes become nearly equal under the same conditions (e.g., front-gate voltage). In addition, front-gate bias allows to shift the resonant tunneling structure, both $R_{XX}$ peaks and dips, relative to the bulk quantum Hall plateau. Experimentally, both peaks and the dips are not affected much by the bulk filling, as in Fig, 8, where the dips continue over two different bulk plateaus and the transition region. Such robustness is puzzling for forward scattering since the precise position of the bulk-constriction edge channel should definitely be affected somewhat by the changing bulk filling, and the tunneling amplitude is exponentially sensitive to the tunneling distance.

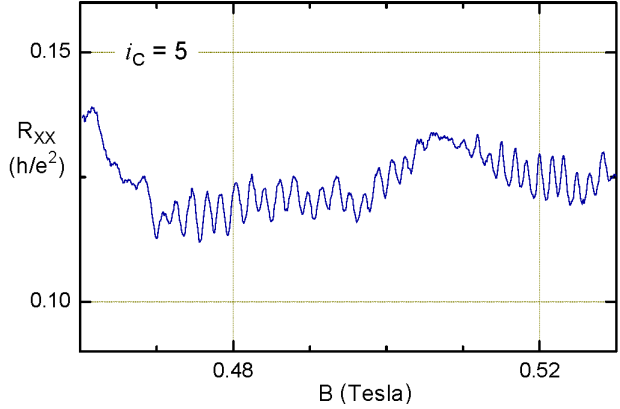

FIG. 9. Resonant tunneling data for $i = 5$ in the QAD constriction.

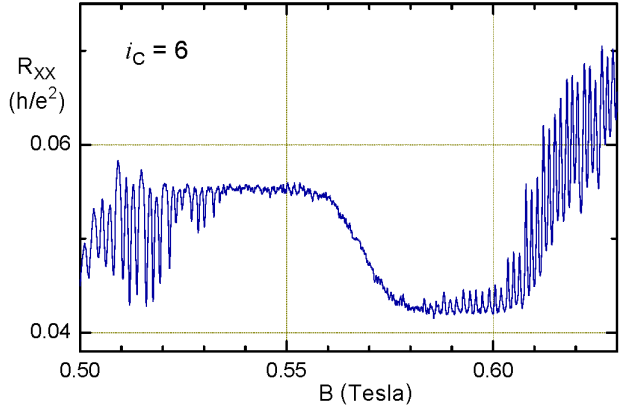

FIG. 10. Resonant tunneling data for $i = 6$ in the QAD constriction. The $R_{XX}$ peaks are seen at $\nu_C < 6$, and the $R_{XX}$ dips for $\nu_C > 6$.

## VI. ANALYSIS AND DISCUSSION

We calculate tunneling conductance from the directly measured $R_{XX}$ vs $B$ data by subtracting the quantized longitudinal resistance and inverting the resulting matrix:

$$G_T = (R_{XX} - R_L)/[(R_H^2 - R_H(R_{XX} - R_L)], \qquad (6)$$

where Hall $R_H = h/i_C e^2$ (since $i_C < i_B$) and longitudinal $R_L = (h/e^2)(1/i_C - 1/i_B)$.[4,11,12,24] Note that both $R_{XX}$ peaks and dips (back- and forward-scattering) result in conductance peaks. Thus obtained tunneling conductance data is plotted in Figs. 11 – 17 and are discussed below.

## A. Magnetic flux period

Figure 11 shows that on the $i$-th integer quantum Hall plateau, the magnetic field interval $\Delta_B \approx 11$ mT containing $i$ tunneling peaks is approximately constant. In other words, the separation between the two neighboring peaks on different plateaus is proportional to $1/i$. This experimental observation leads us to conclude that $\Delta_B$ corresponds to the fundamental antidot flux period $\Delta_\Phi = \Delta_B S_\mu$. This, in turn, allows us to determine the antidot area $S_\mu$ and, assuming a circular antidot, its radius $r_\mu = (h/\pi e \Delta_B)^{1/2} \approx 350$ nm.

Excepting a phase transition,[25] the states $\Psi_{M,\pm}$ of the interacting electrons are adiabatically connected to the corresponding states of the non-interacting system. For non-interacting electrons, the total angular momentum of the electron system $M$ is the sum of the angular momenta $(m)_N$ of occupied orbitals in all $0 \leq N \leq i-1$ spin-polarized Landau levels on the $i$-th plateau. The $M$ increments by one going from one tunneling peak to the next, and, since addition of flux $h/e$ increases number of antidot-bound orbitals by one in each Landau level, there is a total of $i$ tunneling peaks in the fundamental period $\Delta_\Phi = h/e$, as seen in Fig. 11.

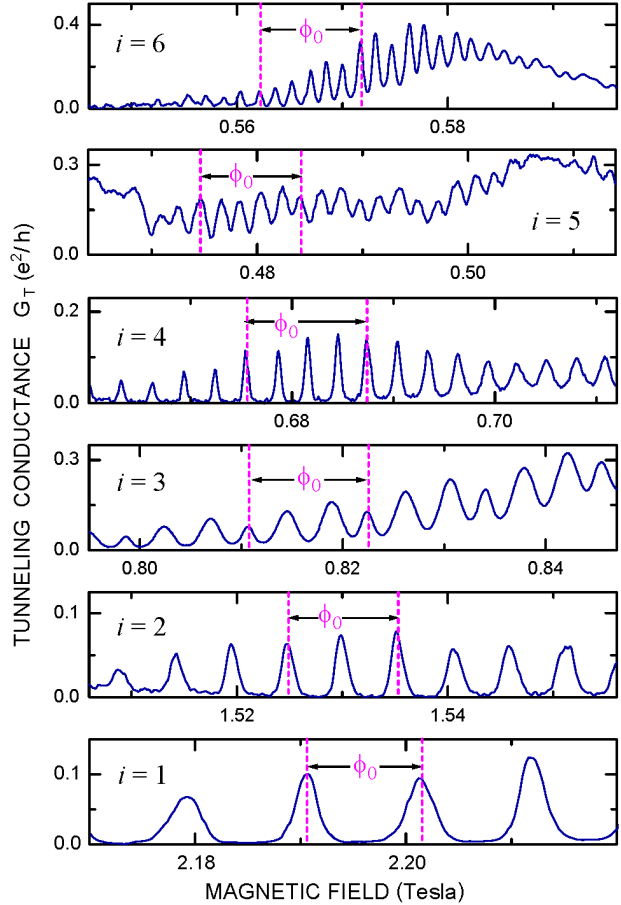

FIG. 11. Representative $G_T$ vs $B$ data for $i = 1$ to 6 in the QAD, plotted on the same magnetic field scale ($\phi_0 \equiv h/e$). The fundamental magnetic flux period $\Delta_\Phi = h/e$ contains $i$ tunneling peaks on the $i$-th integer quantum Hall plateau.

The exact position and relative amplitude of the $i$ peaks within the period $\Delta_\Phi = h/e$ depend on particulars of the antidot-bound electron system. The two limiting cases can be considered. For non-interacting electrons, of tunneling peaks occur at magnetic fields such that the single-particle energy levels $E_{m,N,\pm}$ cross the chemical potential $\mu$. The positions of the peaks then depend on the details of the confining potential, and are not expected to be equally spaced, in general. Since the several Landau levels cross $\mu$ at different positions at the edge, so that the tunneling distance is different for each Landau level, the amplitudes of the $G_T$ peaks within the period $\Delta_\Phi = h/e$ would be expected to exhibit exponentially large variation.

In the other limit, when Coulomb interaction dominates, $e^2 n^{1/2}/4\pi\varepsilon\varepsilon_0 \gg \hbar\omega_C$, the various many-electron ground states $\Psi_{M,\pm}$ within a period have nearly equal occupation amplitudes $c_{m,N,\pm}$, for $m \gg 1$, see Eq. (5). Then the peaks within the period $\Delta_\Phi = h/e$ are expected to be equally spaced, and the amplitudes of the $G_T$ peaks are not expected to exhibit large variation. The experimentally observed $G_T$ vs $B$ data seems to correspond to the strongly interacting limit, as could be expected, because at 1 Tesla the characteristic energies are: $\hbar\omega_C = 1.7$ meV, $\mu_B g^* B = 0.025$ meV, the interelectron Coulomb interaction $e^2 n^{1/2}/4\pi\varepsilon\varepsilon_0 = 3.6$ meV dominates.

### B. Back gate (charge) period

We use the backgate technique[4-6] to directly measure the charge of the QAD-bound particles in the quantum Hall regime. Figures 12, 13, 15 and 16 show $G_T$ vs back-gate voltage $V_{BG}$ data (at a fixed $B$) for antidot on the $i = 1$, 2, 3 and 4 plateaus. All except that for $i = 3$ show approximately equally spaced $G_T$ peaks. The $i = 3$ data show a systematic pairing of the peaks, that is, two alternating peak separations, one consistently less than the other.

Because the global back gate is remote, separated from the 2DES by a $d = 0.430$ mm thick GaAs substrate, the voltage needed to attract one electron to the area of the antidot is large, $V_{BG} \sim 1.5$ V, and the classical electrostatics dominates the small quantum corrections.[13,26] Measuring $\Delta_B$ and $\Delta_{V_{BG}}$, the magnetic field and the back-gate voltage separation of the two *matching* tunneling peaks, allows to determine the particle charge. In this limit, the experimentally-determined magnitude of the charge of the QAD-bound particles is

$$q = \frac{\varepsilon\varepsilon_0}{d} \cdot (h/e) \cdot \frac{\Delta_{V_{BG}}}{\Delta_B}, \tag{7}$$

where $\varepsilon = 12.74$ is the low-temperature GaAs dielectric constant.[27] In practice, accuracy of the measurement is improved by averaging over several (~10) matching $G_T$ peak pairs.

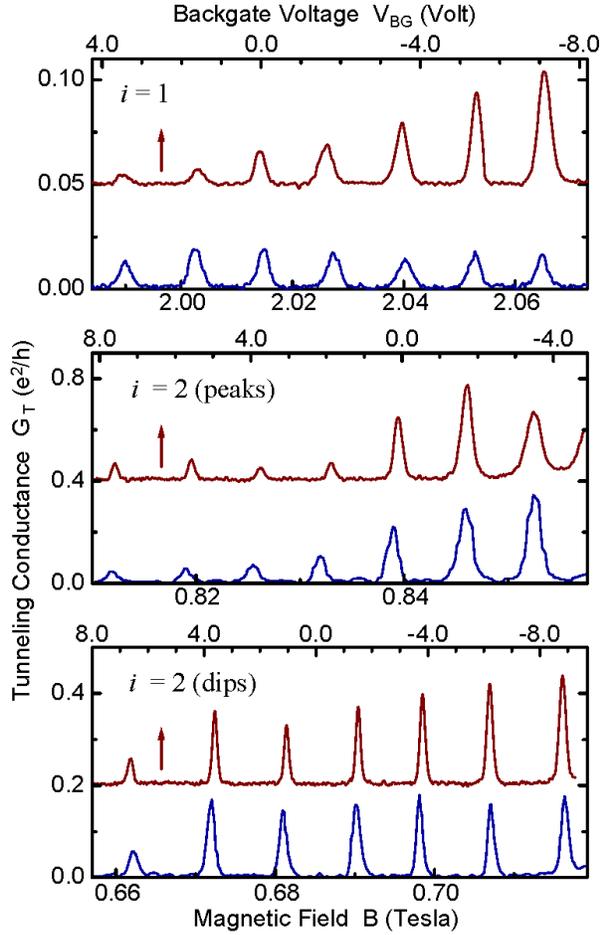

FIG. 12. Tunneling conductance $G_T$ vs magnetic field $B$ at fixed $V_{BG} = 0$ (lower traces, blue) and vs back gate voltage $V_{BG}$ at a fixed $B$ (upper traces, $G_T = 0$ level is shifted up for clarity, red).

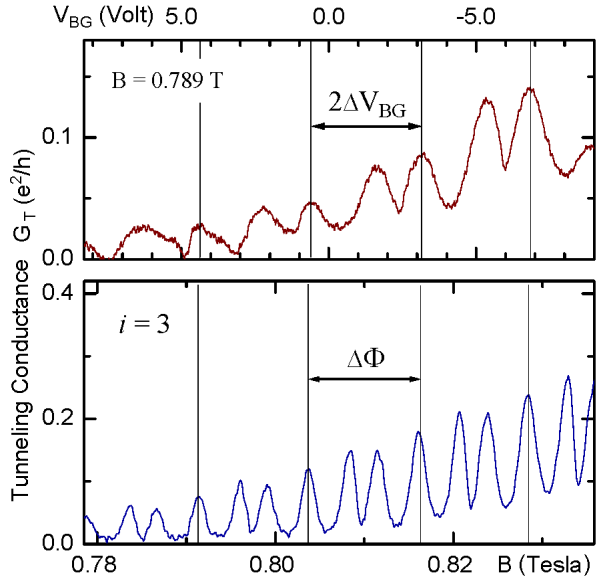

FIG. 13. Tunneling conductance as a function of magnetic field $B$ at $V_{BG} = 0$ (lower panel) and back gate voltage $V_{BG} = 0$ at $B = 0.789$ T (upper panel). The fundamental period $\Delta_\Phi = h/e$ contains three conductance peaks, and the apparent backgate period $2\Delta_{V_{BG}}$ contains two peaks (each $G_T$ peak corresponds to change of QAD occupation by one electron).

The origin of the puzzling internal structure in the $i = 3$ data is not fully understood. If the antidot were much smaller, so that the number of the QAD-bound electrons was small, one would expect non-periodic behavior, like in few-electron quantum dots.[28,29] However, such behavior would be expected to be even more pronounced for $i = 1$ and 2, which is not the case.

Another possibility is the effect of quantum Hall edge reconstruction for $i = 3$, when the spin-split quantum Hall gap is small. The temperature dependence of the conductance peaks at $i = 3$ presented in Fig. 14 shows that all peaks have similar behavior. All conductance peak amplitudes reduce by a factor of two when temperature is raised to ~ 50 mK. This seems to rule out unequal "addition energies" for the QAD-bound states as the origin of the structure. Future numerical modeling of this regime will hopefully elucidate the physical origin of this interesting effect.

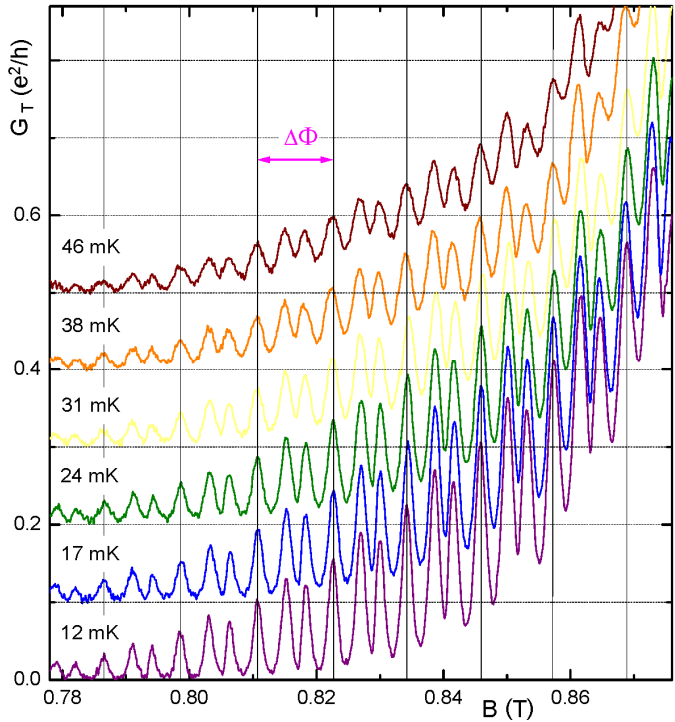

FIG. 14. The temperature dependence of the tunneling conductance peaks at $i = 3$. Note that the three-peak structure within the fundamental period $\Delta_\Phi$ has the same gross temperature dependence as the peak amplitude.

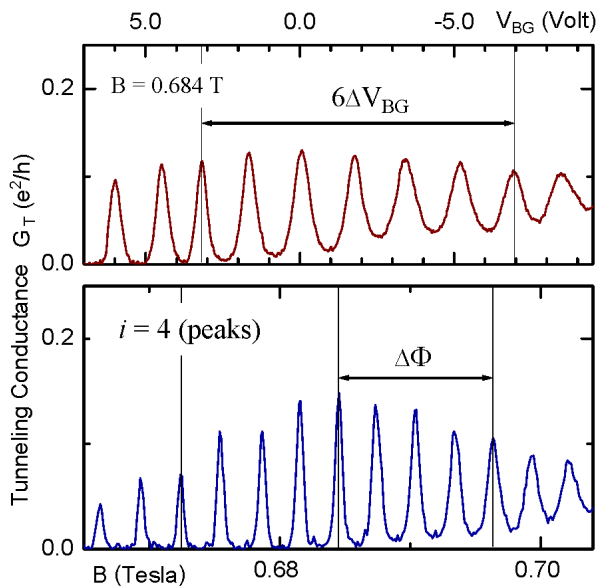

FIG. 15. Tunneling conductance as a function of magnetic field $B$ at $V_{BG} = 0$ (lower panel) and back gate voltage $V_{BG} = 0$ at $B = 0.684$ T (upper panel) for $i = 4$ four-terminal resistance peaks. The fundamental period $\Delta_\Phi = h/e$ contains four conductance peaks, and the period $\Delta_{V_{BG}}$ contains one peak.

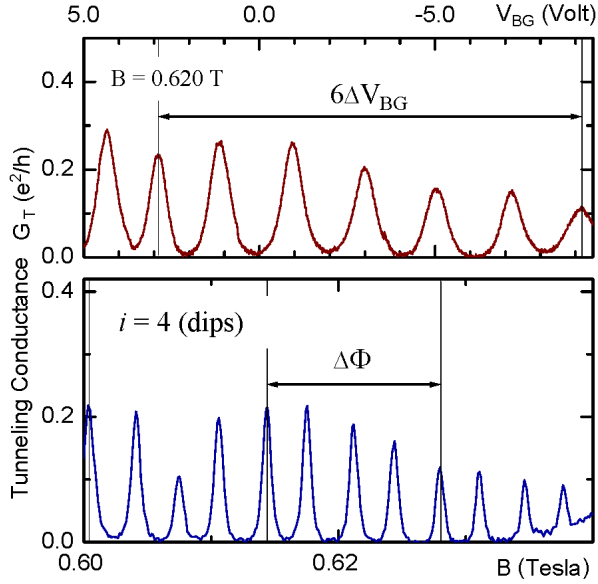

FIG. 16. Tunneling conductance as a function of magnetic field $B$ at $V_{BG} = 0$ (lower panel) and back gate voltage $V_{BG} = 0$ at $B = 0.620$ T (upper panel) for $i = 4$ four-terminal resistance dips. The fundamental period $\Delta_\Phi = h/e$ contains four conductance peaks, and the period $\Delta_{V_{BG}}$ contains one peak.

For the wide $i = 1$ and 2 plateaus, we have investigated dependence of the $\Delta_{V_{BG}} / \Delta_B$ ratio on the position on the plateau, that is, on the filling factor $\nu$. At several $\nu$ on each plateau, we took high-resolution $B$-sweep data fixing $V_{BG} = 0$, then, at several $B$, we took the corresponding $V_{BG}$-sweep data. The periods are determined as the average separation of between six to ten regularly-spaced consecutive $G_T$ peaks, with a "phase slip", "jump" or other irregular data excluded. The results are summarized in Fig. 17, which also shows the tunneling charge $q$ calculated using Eq. 7. The $\Delta_{V_{BG}} / \Delta_B$ ratio remains constant within a standard deviation of 0.7%, which we interpret as evidence that the ratio indeed measures the charge of the antidot-bound particles.

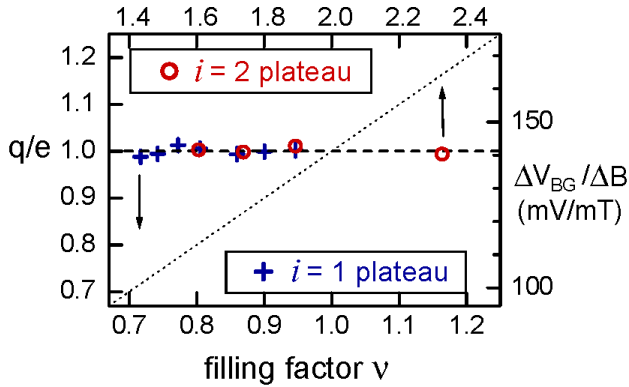

FIG. 17. The experimental period ratio $\Delta_{V_{BG}} / \Delta_B$ as a function of $\nu$ on two quantum Hall plateaus. The dotted line gives the unit slope $\Delta_{V_{BG}} / \Delta_B = \nu / i$. The dashed horizontal line gives constant charge $q = e$. It is evident that the period ratio $\Delta_{V_{BG}} / \Delta_B$ is not proportional to $\nu$, and indeed gives $q$.

An alternative interpretation that the ratio might be proportional to the filling factor is clearly not supported by our experiments. Note that the *position* of a given $G_T$ peak may indeed *approximately* correspond to a fixed $\nu$, so that the peak positions form a $\nu = const$ fan diagram on a $B$ vs. $V_{BG}$ plot. The fan diagram however contains different information than the $\Delta_{V_{BG}} / \Delta_B$ ratio, where the consecutive $G_T$ peak separation (the periods) are compared. These two different measurements would be compatible, *assuming* the $G_T$ peak separation on a given plateau is proportional to $\nu \propto 1/B$, as expected for non-interacting electrons. Such an assumption is not supported by our $B$-sweep data over a wide interval of $\nu$ on a given plateau, for example in Fig. 5 and in Ref. 5; we find a much weaker dependence of the peak spacing on $B$. The accuracy of the reported[30] $B$ vs. $V_{BG}$ fan diagrams is not sufficient to establish whether the peak positions follow straight lines with ($B = 0$, $n = 0$) intercepts of the constant filling $\nu = hn/eB$.

### C. Tunneling peak lineshape and thermal excitation

Figures 12, 13, 15 and 16 show the low-bias, linear resonant tunneling conductance at a low $T$. A high-resolution study of lineshape of a well-separated tunneling peak and its temperature evolution were reported in Ref. 11. Theoretical models consider several limiting regimes, depending on relative magnitude of the tunneling rates $\Gamma_{L,R}$ and the antidot-bound level energy spacing $\Delta E$.[31,32] Any degeneracy in antidot-bound states is assumed to be lifted by the electron-electron interaction, as discussed in Sec. II, consistent with the best experimental results.[11,12] In the intrinsic broadening regime, when $kT \ll \Gamma_{L,R} \ll \Delta E$, so that only one non-degenerate state is involved in tunneling, an isolated, single tunneling peak conductance does not depend on temperature, and has a Lorentzian lineshape:

$$G_T = \frac{e^2}{h} \cdot \frac{\Gamma_L \Gamma_R}{(\mu - E_0)^2 + \Gamma^2}, \tag{8}$$

where $\Gamma = \frac{1}{2}(\Gamma_L + \Gamma_R)$, $\mu$ is the chemical potential, and $E_0$ is the energy of the resonant QAD-bound state via which the tunneling occurs.

In the thermal broadening regime, when $\Gamma \ll kT < \Delta E$ and only one non-degenerate antidot-bound state is involved in tunneling, the tunneling peak lineshape is given by the energy derivative of the Fermi-Dirac distribution:

$$G_T = G_P [\cosh(\frac{\mu - E_0}{2kT})]^{-2}, \tag{9a}$$

$$G_P = \frac{e^2}{h} \frac{\pi \Gamma_L \Gamma_R}{4kT\Gamma}. \tag{9b}$$

In the classical Coulomb blockade regime, when $\Gamma, \Delta E \ll kT$, the tunneling proceeds through many nearly degenerate states, and

$$G_T = G_P \frac{(\mu - E_0)/kT}{\sinh[(\mu - E_0)/kT]}, \tag{10a}$$

$$G_P = \frac{e^2}{h} \frac{\rho \Gamma_L \Gamma_R}{2\Gamma}, \tag{10b}$$

where $\rho$ is the density of the near-degenerate QAD-bound states at $\mu$.

The analysis of the experimental data is described in more detail in Ref. 11. The important conclusions are as follows. Both the peak lineshape and the temperature dependence are consistent with resonant tunneling via one non-degenerate antidot-bound state. As discussed in Sect. II, this is a many-electron ground state of the system. Thermal excitation probes the energy scale of the excited (many-electron) states. The parameter $\alpha = 56.6$ μeV/V gives the "addition energy" level spacing of $\alpha\Delta_{V_{BG}} \approx 150$ μeV for the data of Fig. 19, which can be compared to 30 μeV obtained for the sample of Ref. 13. These energies are sizable fractions of the relevant quantum Hall (tunneling) gaps of ~1.5 meV, nevertheless, the tunneling gap contains many excited antidot-bound states.

These energies can be thought about as the increment of the self-consistent (screened) confining potential at the chemical potential over the distance separating two consecutive basis orbitals, $-(r_{m+1} - r_m)(\partial U/\partial r)$, where $r_m = r_\mu$, as illustrated in Fig. 3(b). In Ref. 21, $\partial U/\partial r$ is calculated for a quantum antidot geometry similar to that in Ref. 10 within a density functional theory in the local spin density approximation. Their results for $i = 2$ in Fig. 2 give $-(r_{m+1} - r_m)(\partial U/\partial r) \approx 90$ μeV at $r_\mu \approx 280$ nm, in a reasonable agreement with experiment.

## VII. CONCLUSIONS

In conclusion, we have experimentally studied electron transport in quantum antidots in the integer quantum Hall regime. In these devices, the antidot-bound electron states are probed by resonant tunneling. On the constriction plateaus $i = 1 - 6$, we find that the tunneling peak spacing is approximately proportional to $1/i$, so that the fundamental flux period $\Delta_\Phi = h/e$ contains $i$ tunneling peaks. The corresponding magnetic field period $\Delta_B$ comprises addition of $i$ single-electron basis states to the antidot area, resulting in $i$ tunneling peaks. The back-gate charging period $\Delta_{V_{BG}}$ corresponds to addition of one electron per tunneling peak within the quantum Hall fluid comprising the antidot-bound electrons. This is interpreted as evidence of the dominance of the electron-electron interaction in the 2DES surrounding the antidot, which mixes the single-particle Landau level occupation. We also analyze the temperature evolution of a well-separated tunneling peak and find the data to be consistent with tunneling through one non-degenerate antidot-bound state.

## ACKNOWLEDGMENTS

We would like to acknowledge participation of Bo Su, Ilari Maasilta and Ismail Karakurt in the earlier stages of this work. This work is supported in part by the National Science Foundation under grant DMR-0555238. The work at Brown was supported by the NSF MRSEC under DMR-0079964 and under DMR-0302222.